\documentclass[11pt,a4paper]{article} 

\usepackage{anysize}
\usepackage[format = hang]{caption}
\usepackage{amsmath,amsthm,verbatim,amssymb,amsfonts,amscd, graphicx}
\usepackage{graphics,natbib}
\usepackage{titlesec,footmisc}
\usepackage{hyperref} 

\usepackage{listings}
\usepackage{booktabs} 

\theoremstyle{plain}

\theoremstyle{definition}

\def\E{{\rm E}}
\def\Var{{\rm Var}}

\def\midd{\,|\,}

\def\dd{{\rm d}}
\def\hatt{\widehat}

\def\bet{{\rm Beta}}

\def\Bin{{\rm Bin}}
\def\something{0.50} 

\def\beq{\begin{eqnarray}}
\def\eeq{\end{eqnarray}}

\def\beqn{\begin{eqnarray*}}  
\def\eeqn{\end{eqnarray*}}

\def\E{{\rm E}}
\def\Var{{\rm Var}}

\def\dd{{\rm d}}

\def\Pr{P}

\def\quadandquad{\quad {\rm and} \quad}

\def\hatt{\widehat}

\def\sumin{\sum_{i=1}^n}

\def\rootn{\sqrt{n}}

\def\midd{\,|\,}

\def\girl{{\rm girl}}

\def\cc{{\rm cc}}

\titleformat{\section}{\normalfont\large\sc\centering}{\thesection}{1em}{}
\titleformat{\subsection}[runin]{\normalfont\large\bfseries}{\thesubsection}{1em}{}
\numberwithin{equation}{section} 
\renewenvironment{abstract}
               {\list{}{\rightmargin\leftmargin}%
                \item[\text{\hspace{10mm}\sc Abstract.}]\relax}
               {\endlist}

\begin{document}

\def\heute{March 2026}

\begingroup
\begin{centering} 

  \Large{\bf Overdispersed and Markovian Children}
  \\[0.8em]
\large{\bf Nils Lid Hjort} \\[0.3em] 
\small {\sc Department of Mathematics, University of Oslo} \\[0.3em]
\small {\sc April 2026\footnote{material partly from a
  FocuStat Blog Post, August 2018; 
  in this modified form April 2026 for wider channels}}\par
\end{centering}
\endgroup



\begin{abstract}
\small{
  Take a look around you -- in your family, your school or workplace,
  in the streets, and you see boys {\it\&} girls in about equal
  proportion, and without any easily visible gender patterns
  in case of siblings. So, to the famous first order of
  statistical approximation, we're all the results of hierarchical
  cascades of independent coin tosses through history, with each
  little fate determined by a 0.50-0.50 coin. This is not entirely
  correct, as one discovers with careful analysis and enough data:
  the coins of fate are (a little) imbalanced;
  they vary (a little) from family to family;
  there is a (slight) dependence in your children's gender sequence;
  and there are (slightly) more only-girls and only-boys
  families than predicted from binomial conditions.
  In this note I use the opportunity to talk also about
  how sample sizes influence p-values and statistical detection power.}

\noindent
{\it Key words:}
beta-binomial,
gender ratio, 
Markovian children, 
overdispersed children
\end{abstract}


\begin{figure}[h]
\centering
\includegraphics[scale=0.33]{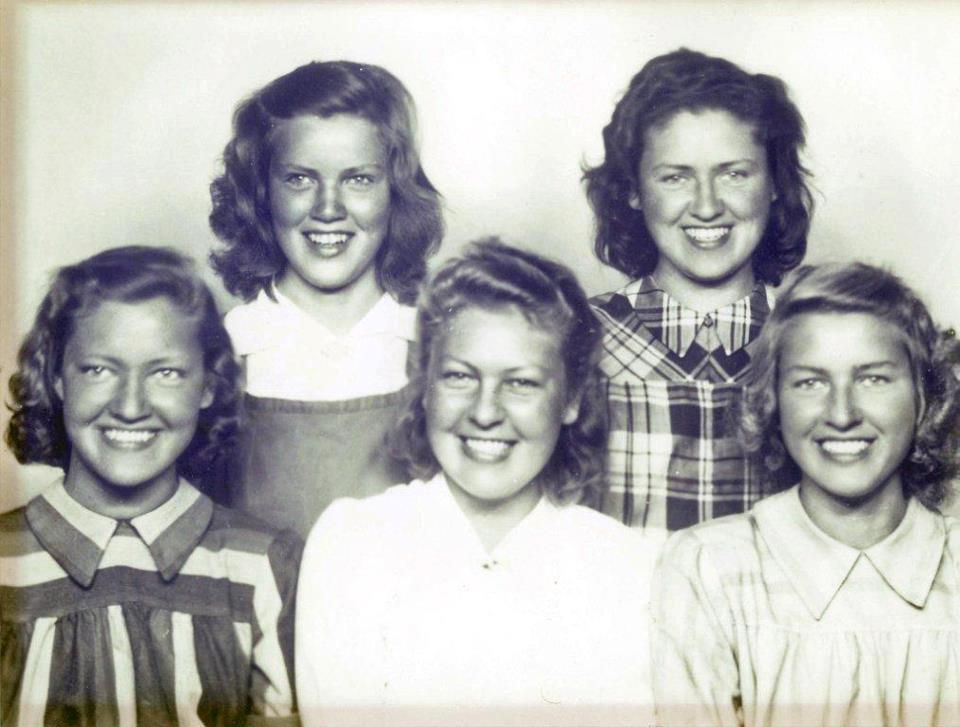}
\caption{$y=5$ girls, $m-y=0$ boys.}
\label{figure:queen1}
\end{figure}

\section*{Introduction}
\label{section:intro}

Once upon a time I saw these numbers in a book in the
statistics library, catching my immediate attention: 
264, 1655, 4948, 8498, 10263, 7603, 3951, 1152, 161.
These relate to the fascinatingly high number $n=38495$
of families in Sachsen, at the end of the 19th century,
who all had at least $m=8$ children, with 264 having 0 girls
(i.e.~8 boys), 1655 1 girl (i.e.~7 boys), etc., up to
161 with 8 girls (i.e.~0 boys). I thought it would be
a perfect example of binomial overdispersion (that there
is more going on than all families having the same binomial
girl-probability $p$), so I jotted down the numbers,
biked home, and wrote up a long exercise for the statistical
inference course I was giving the morning after.
Going through a few calculations one learns
(a) that $p$ is not 0.50, but about 0.485;
(b) that $p$ is likely to vary (not much, but a bit) across families;
(c) that there is enough data to pinpoint these small differences
accurately, and to assess the degree to which $p$ varies.
One also finds a satisfactory explanation of the mildly
surprising statistical fact that there are slightly more
royal straight flush families, with all girls or with all boys,
than it `should be', if the world of children-making
had been binomial with the same $p$. 


Via certain conference talks and conversations and also
some recent literature I decided to write up a
{\it FocuStat Blog Post} about these themes, 
wishing also to use the opportunity to examine and illustrate
how high sample sizes influence both p-values and statistical
detection power. This is more or less what I've put down in
Sections 1-2-3-4 below. But during that evening of work
I found Edwards (1958), with tables of the
Arthur Gei\ss ler Sachsen 1889 data
for  families of sizes $m=2,3,\ldots,12$
(i.e.~not only for $m=8$)\footnote{on another occasion
   I might go into his {\it Über die Phantasmen während des Einschlafens},
   Philosophische Studien, 1878, 1, 83--93},
along with interesting statements
concerning `more boys with bigger families' (see also
Lindsey and Altham, 1998) -- so of course I needed to go through
those tables too, with careful analysis for each. I do confirm
one of these claims, that there is more  overdispersion
with bigger families, but doubt the claim that there will
be more boys.

Playing with these data and models I was also sufficiently
curious to include analyses for Markovian children,
where the gender of your next child depends (well, slightly)
on the gender of your currently last child.
There is a statistical challenge there, as the Gei\ss ler data
only give us the number of families with $y$ girls and $m-y$ boys, 
not the order of genders for children $1,2,\ldots,m$,
but I manage, via simulated log-likelihoods etc.
These extra things make up Sections 5-6 below, and at the end
I have a little list of concluding remarks.

Some of my readers might be content to read through
Sections 1-2-3-4, the basic story about the Sachsen data,
Nature's overdispersion, the beta-binomial, and the role
of sample sizes. Those with stronger appetite may then
paddle through Section 5-6 too, with the Markovian children
and some more advanced statistical tools. Former Eder, I Skabhalse.


\section{Natural Start Assumptions A and B \\ for the construction of genders}

Again, after the proverbial taking a look around you exercise,
we might agree that we, the heaps \& bunches of homo sapiens,
look more or less like Independent Balanced Coin Tosses.
When a new child is born, Natural Start Assumption A
would be that the proportion $p=\Pr(\girl)$ would be 0.50.
There is even something resembling a formal statistical
argument, based on notions of parental expenditure,
formulated by the famous Sir Ronald Fisher (1930),
that creatures with sexual reproduction should have
such a long-term gender balance. We shall see below that
the girl-probability is slightly lower, around $p=0.485$,
with smaller variations from country to country and
from era to era. The ratio of boys to girls at birth is hence
around $0.515/0.485=1.062$ (and in this paper 
I disregard modern ugly attempts at changing this ratio,
say by opting away boys at the pre-birth or even pre-conception
stage). Clearly we cannot expect to detect such a small difference,
between 0.500 and 0.485, until we've met a healthy number
of girls and boys; this is one of the topics below.

Having accepted such a revised version of Natural Start Assumption A,
namely that there is an approximately constant $p=\Pr(\girl)$, 
the Natural Start Assumption B would be that of statistical
independence: in a family with $m$ children, the number of girls
should be a $\Bin(m,p)$, the classical binomial. In other words
(and symbols), if we line up say 1000 families with $m$ children, 
along the equator, the proportion of those with $y$ girls 
ought to be
\beqn
f(y,p)={m\choose y}p^y(1-p)^{m-y}\quad {\rm for\ }y=0,1,\ldots,m.
\eeqn 
In particular, we should see about $1000\,p^m$
families with only girls and about $1000\,(1-p)^m$ 
with only boys. It turns out that this binomial view of the
world's families provides a reasonable fit to what
one actually observes, particularly for small families
with say $m=2,3,4$, but that this Natural Start Assumption B
doesn't hold up to careful scrutiny either, once enough
data about somewhat bigger families are entered into
the analysis. This is also returned to below, but first
I look into the statistical task of demonstrating that
$p=\Pr(\girl)$ is indeed not 0.50.

\section{How to spot the difference between 0.500 and 0.485}

Suppose we meet as many as $n$ children and note their genders,
with $z$ of them being girls. For simplicity I assume these
come from different families, so that the count variable $z$ 
can be seen as a grand binomial $\Bin(n,p)$.
The overall estimate of $p=\Pr(\girl)$ is $\hatt p=z/n$, 
and a 95 \% confidence interval is
$\hatt p\pm1.96\,\{\hatt p(1-\hatt p)/n\}^{1/2}$. 
If we wish such an interval of confidence to be small enough
to not touch 0.50, if $\hatt p$ should happen to be 0.485,
then a rough but informative translation is that we need
$2\cdot\sqrt{0.50\cdot0.50/n}\le 0.015$, which means
$\rootn\ge1/0.015$, or $n\ge4445$. Slightly more carefully,
we ought to have
\beqn
1.96\,\sqrt{0.485\cdot0.515/0.015}\le\rootn,
\quad {\rm or\ }n\ge 65.304^2.
\eeqn 
So if a patient school teacher or an efficient birth registry
machine bothers to catalogue $n\ge 4265$ children,
and if she or he or it indeed finds $\hatt p=0.485$,
signs are clear that the real $p$ will be below Fisher's envisaged 0.50.

This is not quite enough for the sufficiently serious and
professionally skeptical statistician, however. If the true $p$
is 0.485, the confidence interval could still touch 0.50,
even with 4000 children volunteering for gender sampling,
by the same argument as above. The question raised calls
for a proper statistical test of the null hypothesis
$H_0$ that $p=0.50$ -- and for even bigger sample sizes
to be verifiably effective; you need to toss a 0.485 coin
a lot of times to be very sure that it is not a fair coin.
Building a test is easy textbook material,
and a natural test statistic (among several close relatives) is
\beqn
V_n={z-np_0\over \sqrt{np_0q_0}}, 
\eeqn 
on this occasion with $p_0=0.50$ and $q_0=1-p_0=0.50$.
This test statistic is very nearly a standard normal
under the null hypothesis, and with significance level
the customary 0.05, we reject $H_0$ if $|V_n|\ge1.96$.

The {\it power} of the test is its ability to detect
that the null is wrong, as a function of how wrong it is.
At alternative position $p$, we may write $z$ as
$np+\sqrt{npq}\,N_n$, with $q=1-p$, and hence
\beqn
V_n=pN_n + \rootn{p-p_0\over \sqrt{p_0q_0}}, 
\eeqn 
where $\rho=\sqrt{ pq/(p_0q_0) }$ and $N_n$ is very close
to a standard normal. This makes it an easy enough task
to compute the power function
\beqn
\pi_n(p)=\Pr(|V_n|\ge1.96\midd p)
\eeqn 
as a function of $p$. This power function is plotted
in Figure \ref{figure:queen2} below,
for sample sizes $n=10000$ (black curve)
and $n=15000$ (red curve). At the null value $p=0.50$, 
the power is the chosen significance level 0.05, by construction.
To achieve detection power $\pi_n(p)\ge0.95$,
at position $p$ equal to 0.485 or for that matter 0.515,
it turns out that we need as many children on board as $n\ge14433$. 

\begin{figure}[h]
\centering
\includegraphics[scale=\something]{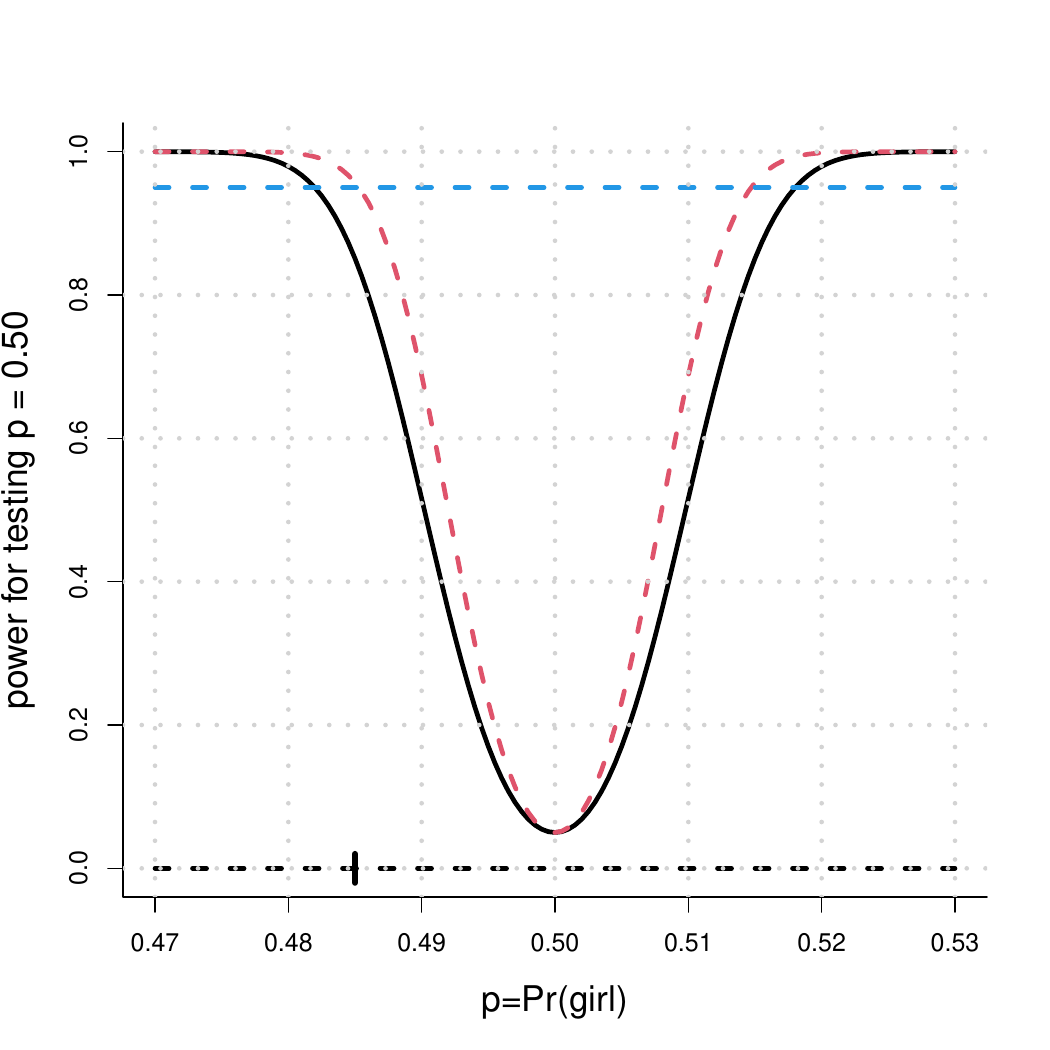} 
\caption{
The power function, the probability of claiming that the true $p$
is not equal to $p_0=0.50$, as a function of $p$, 
plotted for $n=10000$ (black curve) and for $n=15000$
(red curve). At position $p_0$, the power is 0.05,
the chosen significance level of the test, by construction.}
\label{figure:queen2}
\end{figure}

It is a remarkable discovery, and a remarkable achievement,
to be able to claim with high statistical accuracy that human
gender balance is skewed. To illustrate the basic concepts
and tools I chose above the too-often-used significance
level 0.05 and also too-often-used 0.95 certainty level.
With stricter thresholds of accuracy, reflected in
say significance level 0.01 and detection power 0.99,
the required sample size is as high as $n\ge26691$, 
a higher number of children than most of us ever see
in a lifetime. With modern birth registry data this is however
within easy reach. We ought to admire the earliest statistical
efforts that contributed to uncovering this subtle and
gentle bias of God's coins; see e.g.~the discussion
in Stigler (1986, pp.~225-226). Important names here
are John Arbuthnot (in {\it An argument for Divine Provinence,
taken from the constant regularity observ'd in the births
of both sexes}, 1710, he went through 82 years of christening
statistics for London, and formulated his argument
in terms resembling statistical testing and p-values),
and Pierre-Simon Laplace (who in 1778 considered combined
data for half a million births, again providing what
we in modern terms call a p-value).

\section{Children are extrabinomially overdispersed} 

The medical doctor Arthur Gei\ss ler (1832--1902,
inside the reign of Königreich Sachsen, 1806--1918,
making him perhaps a Saxon rather than a German,
though Sachsen became part of the Deutsches Reich in 1871)
was an eager collector of family and birth registry data,
from all over Sachsen. Those were also the times of families
with many children.

\begin{figure}[h]
\centering
\includegraphics[scale=0.60]{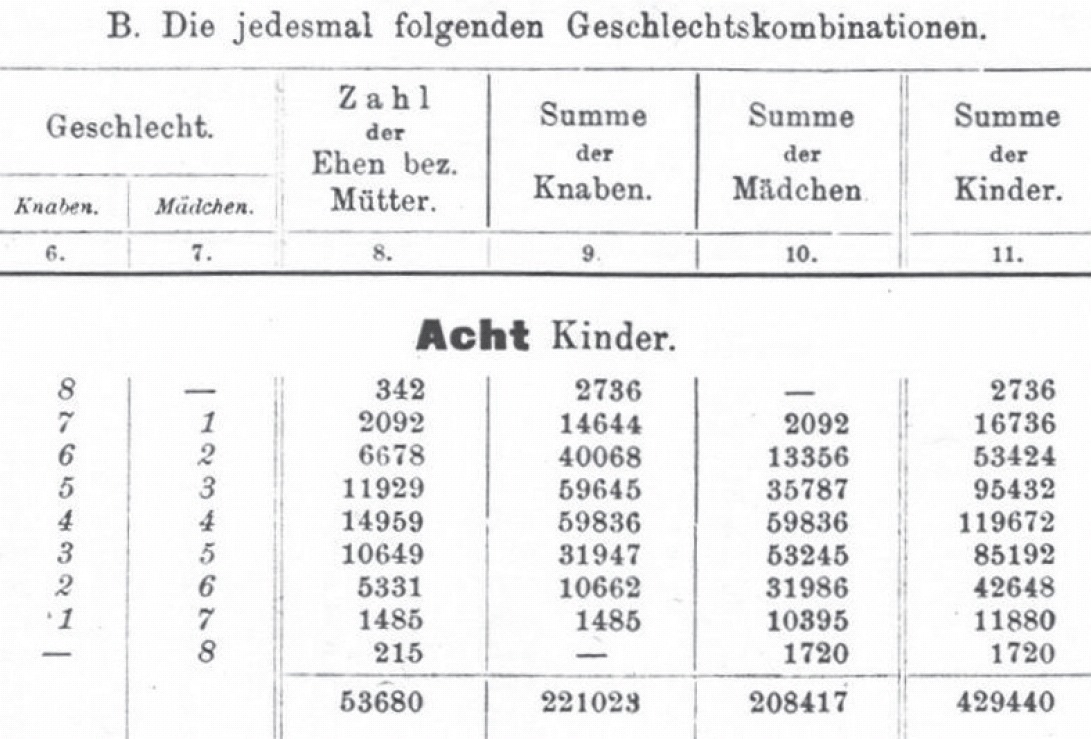}
\caption{One of the tables in Gei\ss ler (1889),
concerning families with eight children.}
\label{figure:queen3}
\end{figure}


From his collections of data tables, R.A.~Fisher and others
have extracted information on the number for girls,
among the first $m=8$ children, for as many as $n=38495$
families, as follows, cf.~Edwards (1958, 2005).
So 264 had 0 girls, 1655 had 1 girl, 4948 had 2 girls, etc.,
see the first two columns in this table:

{{\baselineskip13pt
\begin{verbatim}
             binomial            beta-binomial      Markov
   y   N(y)     E1(y)    pears1     E2(y)   pears2     E3(y)    pears3 
   0    264    192.32     5.17     255.54     0.53    255.96     0.50  
   1   1655   1445.38     5.51    1656.79    -0.04   1654.75     0.01  
   2   4948   4752.36     2.84    4909.50     0.55   4906.78     0.59  
   3   8498   8928.90    -4.56    8683.46    -1.99   8685.61    -2.01  
   4  10263  10484.95    -2.17   10025.35     2.37  10034.41     2.28  
   5   7603   7879.79    -3.12    7736.16    -1.51   7734.77    -1.50  
   6   3951   3701.20     4.11    3896.34     0.88   3893.65     0.92  
   7   1152    993.42     5.03    1171.05    -0.56   1168.12    -0.47  
   8    161    116.65     4.11     160.81     0.01    160.98    -0.00  
      38495  38495      159.41   38495       13.55  38495       13.17        
\end{verbatim} 
}}

\begin{small}
\noindent 
The table: The Gei\ss ler (1889) data, for as many as $n=38495$ 
families with $m=8$ children (or more), with $N(y)$ 
the number of these with $y$ girls, for $y=0,1,\ldots,8$. 
The table then gives the theoretical or expected number $E(y)$,
for models
(i) the binomial,
(ii) the beta-binomial,
(iii) the Markovian, see below,
along with what I call Pearson residuals $(N-E)/\sqrt{E}$.
The smaller these are, in absolute size, the better is the
model behind them. The numbers 159.41, 13.55, 13.17
are the goodness-of-fit statistics, the sum of the squared
Pearson residuals.
\end{small}

\medskip
So, out of $mn=307928$ children, $\sum_{y=0}^8 N(y)=149158$ 
were girls (with $N(y)$ denoting the number of families with
$y$ girls), leading to the overall estimate $\hatt p=0.484$
for the girl-probability. The third column gives the expected number
\beqn
E_1(y)=nf(y,\hatt p)=n{m\choose y}\hatt p^y(1-\hatt p)^{m-y} 
\eeqn
of the $n$ families with $y$ girls, for $y=0,1,\ldots,8$,
under binomial circumstances. The fit is not so bad,
though not in the category of `good',
as revealed by what I term Pearson residuals
\beqn
P_1(y)={N(y)-E_1(y)\over \sqrt{E_1(y)}} 
\eeqn 
for 
for $y=0,1,\ldots,8$ (placed in the ${\rm pearson}_1$ column).
Under the conditions of the model, that there
is a binomial mechanism governing the number of girls,
with the same $p$, these Pearson residuals would
be approximately standard normally distributed,
and with values outside the range $(-2.5,2.5)$
signalling that the fit is not good. Also, under model
conditions, the goodness-of-fit statistic
$Z_1=\sum_{y=0}^8 P_1(y)^2$ should have a $\chi^2_7$
distribution (a chi-squared with degrees of freedom equal to 7,
the number of cells minus 1 minus another 1 for the number
of parameters estimated); here $Z_1=159.82$, 
which is far too big.

So something is rotten in the state of Binomialia:
the terrain has more weight at the borders ($y=0,1,7,8$)
and less near the middle, compared to what the
theoretical map predicts. The real world of the real
children thus exhibits more variability than what
a single binomial $p$ can explain. To shed light
on this overdispersion, assume now that each family
has its own girl-probability $p$
(whether this is due to the mother or the father,
or to their con-divine conjunction habits),
but that these vary from family to family,
following some distribution $g(p)$, say.
For each family we then have the well-known formulae
\beqn
\E\,(y\midd p)=mp \quadandquad \Var\,(y\midd p)=mp(1-p)
\eeqn 
for the mean and variance. Let further $p_0$ and $\sigma_0$ 
be the mean and standard deviation of the background distribution
of $p$ in the population of families. Then, using
the laws of total mean and total variance
(cf.~the double expectation formula), we find
\beqn
\E\,y=mp_0 \quadandquad \Var\,y=mp_0(1-p_0) + m(m-1)\sigma_0^2
\eeqn 
for the mean and variance of what Gei\ss ler and we actually
observe in the population. So $m(m-1)\sigma_0^2$
is the extrabinomial variation, with a small $\sigma_0$ 
corresponding to the distribution of $p$
being tight around its mean $p_0$. Note that with $m=1$, 
a land of one-child families, there is no such
extrabinomial variation.

So let us assess the extra variation. Estimating $\sigma_0$ 
is not entirely straightforward, since it is the standard
deviation of a variable we never actually observe,
only indirectly, via the girl-counts $y_i$
themselves. We may however estimate the variance of the $y_i$
in the usual fashion, via
\beqn
S_n^2={1\over n-1}\sumin (y_i-\bar y)^2
    ={1\over n-1}\sum_{y=0}^m N(y)(y-\bar y)^2, 
\eeqn 
with $\bar y=(1/n)\sum_{y=0}^m N(y)y=3.875$
the overall average number of girls, associated also
with $\hatt p_0=\bar y/m=0.484$, and then subtract
the binomial variance part, i.e.~$m\hatt p_0(1-\hatt p_0)$.
This difference, which turns out to become $2.1605-1.9981=0.1624$, 
is an estimate of $m(m-1)\sigma_0^2$. Dividing with $8\cdot7$ 
and taking the square root yields $\hatt\sigma_0=0.0539$
indicating that about 95 \% of the families of the world
(at least in Sachsen, in the late 19th century) have their $p$
inside $[0.38,0.58]$ -- I provide a more accurate assessment below.

We ought to provide a clear statistical test for
the presence of overdispersion. A natural test statistic is
\beqn
W_n={S_n^2\over m\hatt p_0(1-\hatt p_0)}, 
\eeqn 
since this ratio will be close to 1 under fixed $p$ 
circumstances but bigger than 1 in the presence
of a positive $\sigma_0$. Here we find $W_n=1.081$, 
which might not look impressive, but is;
the 0.01 and 0.99 quantiles of the null distribution of $W_n$, 
based on $10^4$ simulations under binomial conditions,
are 0.984 and 1.016.

\begin{figure}[h]
\centering
\includegraphics[scale=\something]{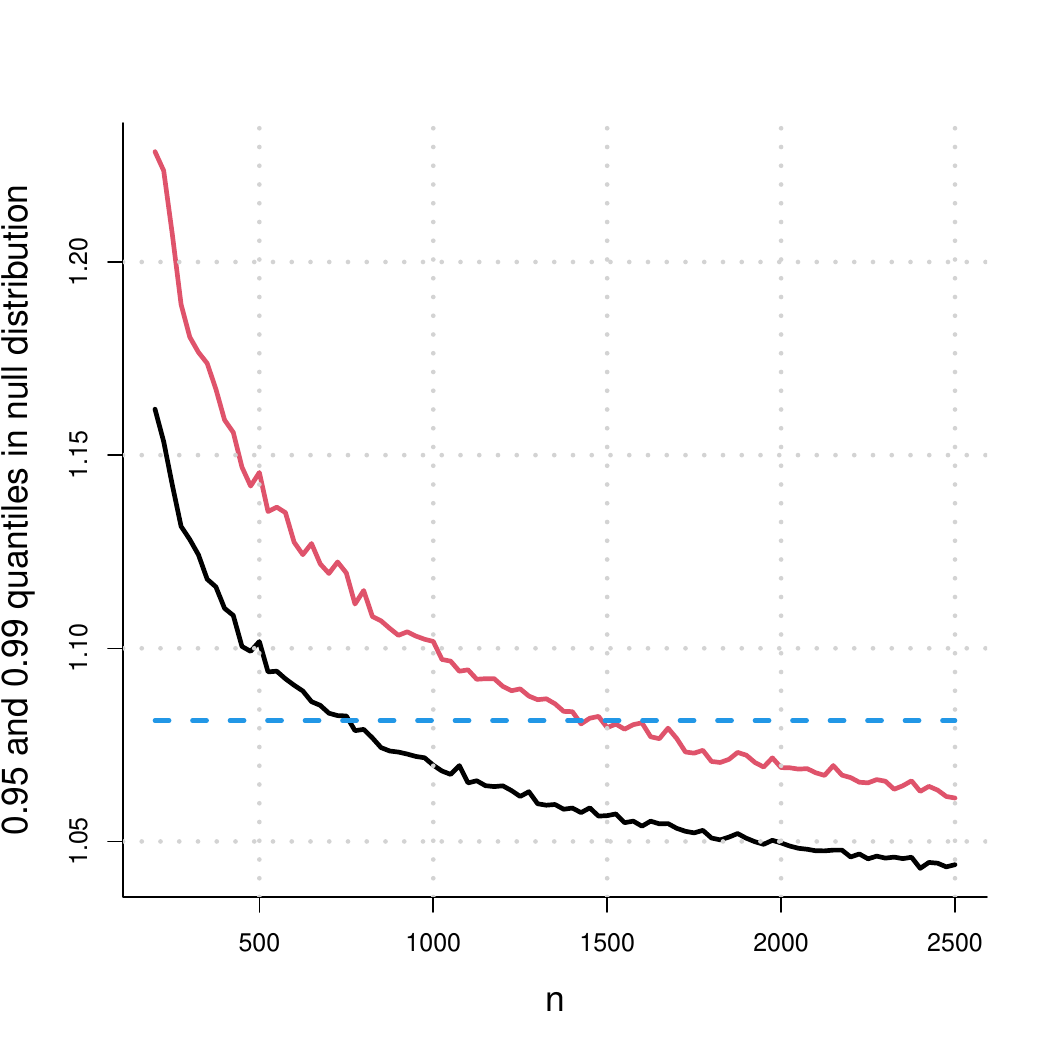}
\caption{The 0.95 and 0.99 quantiles (black and red)
  for the null distribution of $W_n$, as a function of
  sample size $n$.}
\label{figure:queen4}
\end{figure}

I pause briefly to comment on how the sample size
influences both p-values and detection power. Observing
the full-variance to under-model-variance ratio $W_n=1.081$
is found to be superconvincing evidence for overdispersion
here, due to $n$ being so large; with a smaller $n$
it might not qualify as significant. Figure \ref{figure:queen4}
displays the 0.95 quantile (black) and 0.99 quantile (red)
in the null distribution of $W_n$, as a function of sample size.
So the observed value 1.081 starts getting interesting
at around $n=750$ (p-value below 0.05)
and very convincing at around $n=1500$
(p-value below 0.01); and, of course, with $n=35000$
or more, as with the Sachsen data, the null hypothesis
associated with the pure binomial is blown away.

\section{A better model: the beta-binomial}

A very reasonable model to try out is to place a $\bet(a,b)$ 
distribution on the $p$, with density say $g(p,a,b)$.
The distribution of the girl-count $y$ in a family of $m$
children then takes the form
\beqn
f_2(y,a,b)
&=&\int_0^1 f(y,p)g(p,a,b)\,\dd p \\
&=&\int_0^1 {m\choose y}p^y (1-p)^{m-y} {\Gamma(a+b)\over \Gamma(a)\Gamma(b)}
p^{a-1}(1-p)^{b-1}\,\dd p \\
&=& {m\choose y}{\Gamma(a+b)\over \Gamma(a)\Gamma(b)}
   {\Gamma(a+y)\Gamma(b+m-y)\over \Gamma(a+b+m)}, 
\eeqn
for $y=0,1,\ldots,m$. The parameters $(a,b)$ can be fitted
via maximum likelihood, by minimum chi-squared,
or by equating the observed mean $\bar y$ and variance $S^2$ 
to the corresponding population quantities
\beqn
\E\,y=mp_0=m{a\over a+b} 
\eeqn 
and
\beqn
\Var\,y=mp_0(1-p_0) + m(m-1)\sigma_0^2
   =mp_0(1-p_0) + m(m-1){p_0(1-p_0)\over a+b+1}. 
\eeqn 

This moment estimation method leads directly to
$\hatt p_0=y/m=0.484=a/(a+b)$ again, and a simple equation
for finding $a+b$. 

\begin{figure}[h]
\centering
\includegraphics[scale=\something]{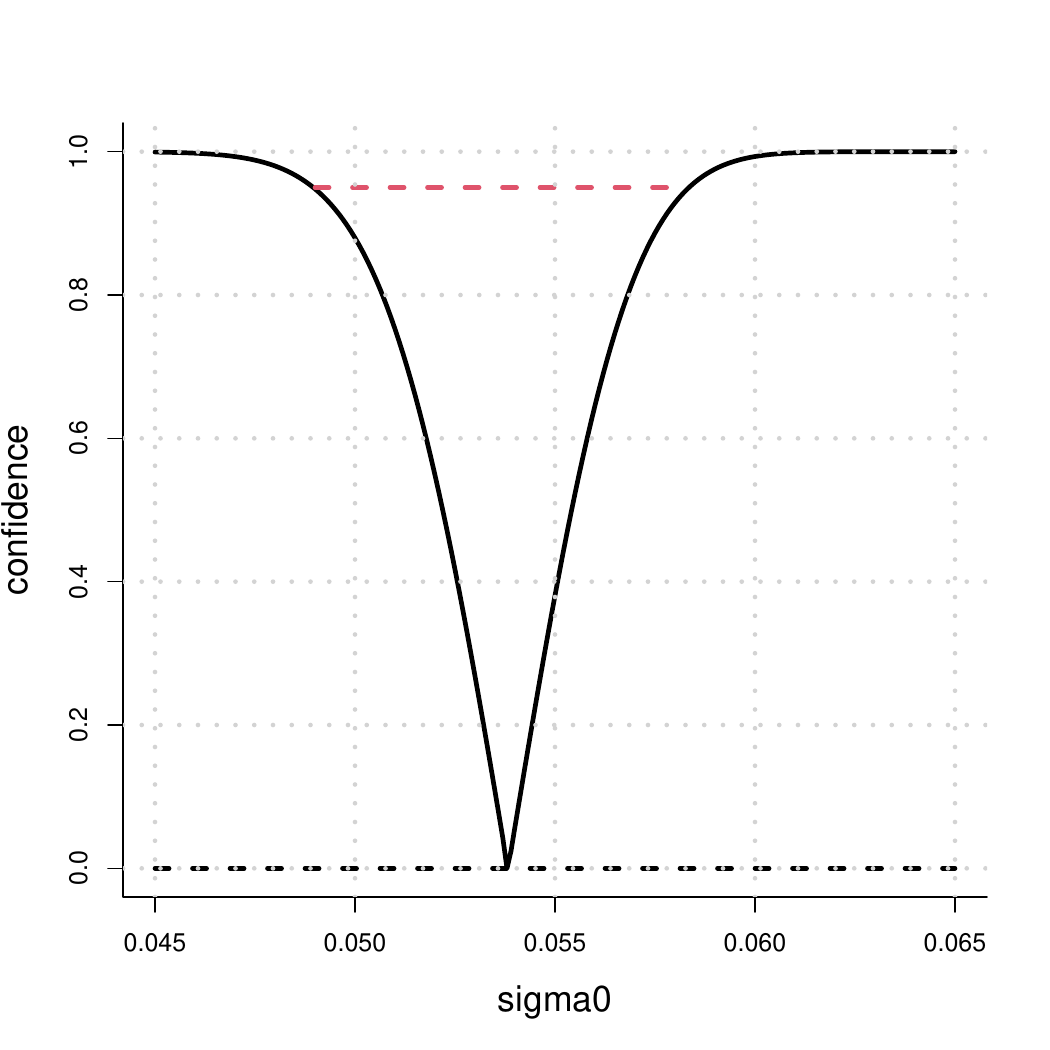} 
\caption{The confidence curve $\cc(\sigma_0)$ for the overdispersion
  parameter $\sigma_0$; the point estimate is 0.0538
  and the 95 \% interval is $[0.0490,0.0583]$.
  Pure binomial, the same $p$, means $\sigma_0=0$.}
\label{figure:queen5}
\end{figure}

These three estimation approaches give rather similar results
for the 8-children families from 19th century Sachsen.
The minimum chi-squared method consists in minimising
the sum of the Pearson residuals
\beqn
Q_n(a,b)=\sum_{y=0}^8 P_2(y)^2
   =\sum_{y=0}^8 { \{N(y)-E_2(y,a,b)\}^2\over E_2(y,a,b)}
\eeqn 
over all $(a,b)$, whee $E_2(y,a,b)=nf_2(y,a,b)$ is the expected
numer of families with $y$ girls, under the model being studied.
I find $(\hatt a,\hatt b)=(41.179,43.833)$,
amounting to a Beta density with mean 0.484 and standard deviation
$\sigma_0=0.0538$.

So 95 \% of all couples have their $p$ inside the interval
$[0.396,0.573]$. That the fit of the beta-binomial is very good
is seen via the table above, with fits $E_2(y,\hatt a,\hatt b)$
close to the observed counts $N(y)$, with Pearson residuals
$P_2(y)$ within the standard normal range, in total
offering a drastic improvement on the simple binomial
fixed-$p$ explanation. The minimum sum of squares
$Z_2=Q_n(\hatt a,\hatt b)=13.721$ is admittedly not
entirely perfect, compared to the scale of the $\chi^2_6$ 
(now with degreees of freedom equal to 6, namely the number
of cells minus 1 minus 2 more for the two estimated
parameters), but taking the enormous Saxonic sample size of $n=38495$
into account the statistician really can't ask for a better fit.

The size of the extrabinomial dispersion is of particular interest,
i.e.~the standard deviation $\sigma_0$ in the distribution of $p$
across a broad population of families. I've gone to the trouble
of computing a confidence curve $\cc(\sigma_0)$ for this
parameter, using methods of Schweder and Hjort (2017, Chs. 3-4);
see Figure \ref{figure:queen5}. 

\section{Markovian children}

For the Ge\ss ler data we do not have information of the order
of girls and boys, just their totals. A fascinating data set
analysed in Kotz (1972) has however order information,
for each of a high number of Amish and Mennonite families,
with parents born before 1910. This is from the era where birth
control was considered sinful and with `be fruitful and multiply'
ruling. He fitted these binary sequences to a Markov model,
estimating a certain association parameter, and found this
to be just visible, as in barely significant. In connection
with writing up this article I have accessed these data
and worked with them myself (where a serious and
time-consuming challenge, for me, was to translate the
octogonal coding of Kotz's data back to understandable
binary sequences). I have also confirmed that the Markov
dependence, in the model he set up, is just about visible
for the data, but not in a prominent fashion.

I was curious enough to fit also the Gei\ss ler data to a
natural Markovian model. This takes further efforts,
since I only know that you have $y$ girls and $m-y$ boys 
not the order of your $m$ children. Hence I cannot use
the machinery of Kotz (1972, 1973), but as long as I put
up a parametric model for $x_1,\ldots,x_m$, with $x_i=1$ for
girls and $x_i=0$ for boys, I can for any $\theta$ 
in that model simulate a large number of gender paths
$x_1,\ldots,x_m$, read off the number of girls $y=\sum_{i=1}^m x_i$,
and estimate say $f_3(y,\theta)$, the implied distribution,
by checking the proportion of paths which led to $y$.
In particular, I can compute (well, actually, simulate
good enough approximations to) and work with the log-likelihood function
\beqn
\ell(\theta)=\sumin \log f_3(y_i,\theta)
   =\sum_{y=0}^m N(y)\log f_3(y,\theta), 
\eeqn 
even when there is no explicit formula for $f_3(y,\theta)$. 
This the approach of simulated log-likelihood, obtaining
an estimate $\hatt\ell(\theta)$ through simulation,
for a grid of $\theta$ values.


\begin{figure}[h]
\centering
\includegraphics[scale=\something]{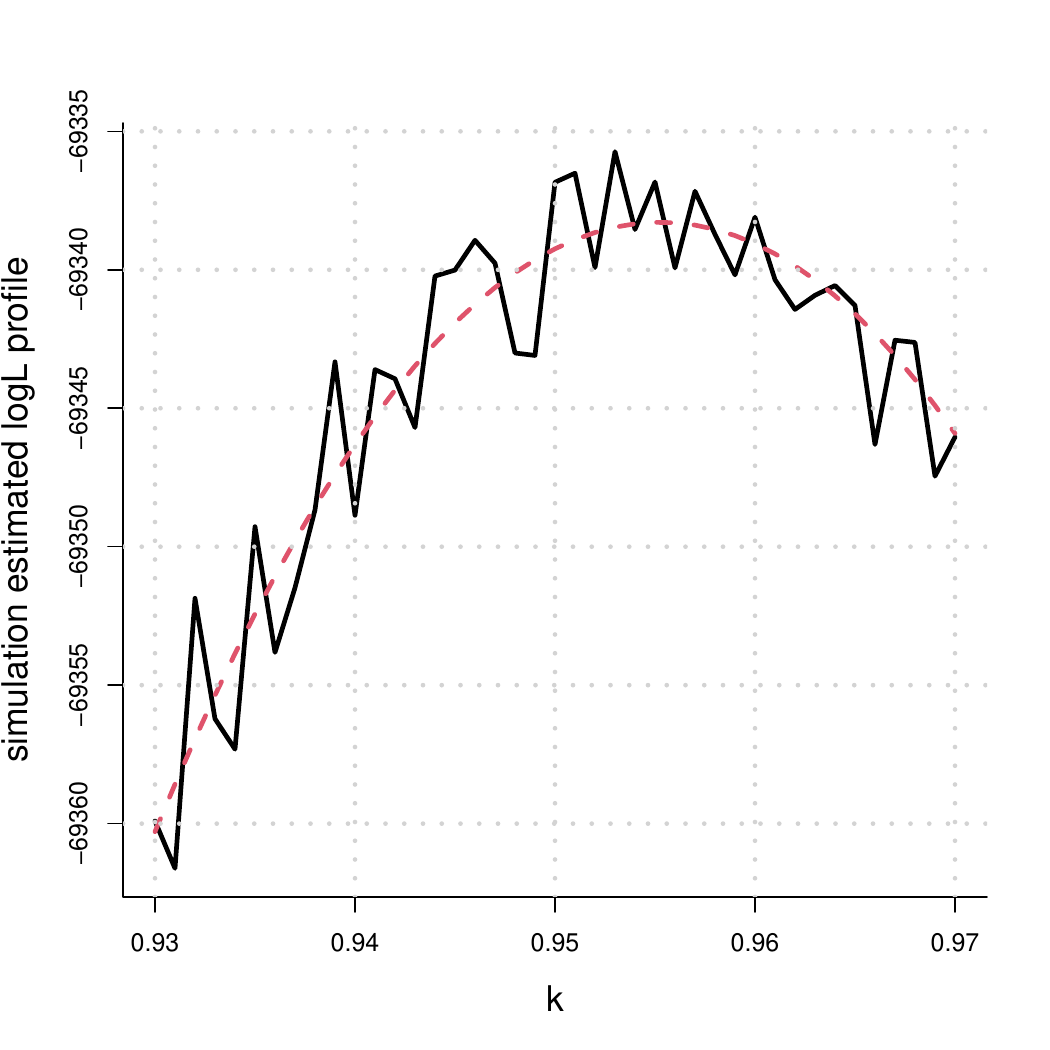} 
\caption{Simulated log-likelihood $\ell^*(k)$, with values
  obtained at a grid of $k$ values, using $10^5$ simulations
  for each, followed by a 4th order polymomial approximation
  to compute the maximiser.}
\label{figure:queen6}
\end{figure}

Let me now introduce my model for the Markovian children,
where the gender of your next child depends (but only slightly,
as it turns out) on the gender of your currently last child.
Let $(q_0,p_0)=(1-p_0,p_0)$ be the long-term frequencies
for boys and girls. Instead of taking births to be fully
independent of each other, consider the Markov chain
$x_1,x_2,\ldots$ of births (again with 0 for boy and 1 for girl),
with $x_1$ drawn from the $(q_0,p_0)$ coin of things,
and then following the two-stage transition probability matrix
\beqn
P=\begin{pmatrix} 1-kp_0 & kp_0 \\ kq_0 & 1-kq_0 \end{pmatrix}, 
\eeqn
with $k$ a fine-tuning parameter. If $k=1$ we're back to
ordinary independence, with $(q_0,p_0)$ for each new birth.
This Markov chain has $(q_0,p_0)$ as its equilibrium
distribution. The covariance beetween consecutive
`it's a girl' and `it's a girl' is
\beqn
p_0(1-kq_0)-p_0^2 = p_0q_0(1-k),
\eeqn 
so the correlation between girl, girl is $1-k$;
the same expression is also valid for the correlation
between boy, boy. We shall find $k$ around 0.95 below,
which then means `same gender twice' correlation around 0.05.

With considerable efforts I have fitted this two-parameter
Markov model to the Gei\ss ler data, for the case of $m=8$ 
as well as for the other family sizes (see below for these data).
This involves somewhat heavy simulations, constructing
figures as the one below, zooming in via the right grids, etc.
For my favourite case of $m=8$, I find I find $\hatt k=0.956$
with 99 \% interval $[0.947,0.964]$, 
which means correlation between (girl, girl),
or between (boy, boy), estimated at 0.044 with interval
$[0.036,0.053]$. 

The fit of this after all relatively simple Markov model
is remarkably good, as seen in the two last columns
of the table in Section 3. The view taken for this model
is that the same parameters $(k,p_0)=(0.956,0.484)$ 
are at work across all families, with transition matrix
\beqn
\hatt P=\begin{pmatrix} 0.537 & 0.463 \\ 0.492 & 0.508 \end{pmatrix}
\eeqn
governing the next child's gender. The minimum chi-squared
statistics is 13.17 (found via an embarrassingly high number
of simulations, to make it precise enough), just a bit smaller
than than for the beta-binomial. I've also gone to the
considerable trouble of computing the maximum log-likelihood
value (again, via simulations) for the Markov model,
comparing this with the corresponding value for the
beta-binomial model, for each of the family sizes $m=2,3,\ldots,12$.
These numbers are needed for model comparisons and
ranking via the model selection criteria AIC or BIC
(see Claeskens and Hjort, 2008). The two models are quite
comparable. Focused categorical model selection via
FIC methods in Hjort and Jullum (2018) may also be brought
to the table, when special questions, like estimating
the probability of all-girls families, are under consideration. 

One may also extend the initial Markov model above
a bit further, to improve the fit a tad or two more,
e.g.~by allowing a distribution for the $p_0$
while keeping the Markov parameter $k$ fixed.
This can be done using the methods developed above,
but with added efforts (model building, programming,
interactive simulations).

\section{Are the sex ratio and the overdispersion level \\
  changing with family size?}

I have excerpted the following table of information from
related data summaries in
Edwards (1958)\footnote{I'm glad to note that I in March 2026
   have been in touch with the author of that 1958 paper,
   regarding details of the present article, 
   and also to hear that he at age 90 has just published
   {\it THE LATIN SQUARE: Essays in Defence of R.A.~FISHER}}
and Lindsey and Altman (1998), who again have used the Sachsen 1889 tables to
provide counts of $y$ girls in families of $m$ children, for
$m=1,2,\ldots,12$. So there were 6115 families with 12 (or more)
children, with the girl probability estimated to
$\hatt p = 35280/(12\cdot6115)=0.481$ for these, etc.

\begin{small}
\smallskip\noindent 
Table: giving the number of Sachsen families, sorted by
family size, with a given number of girls. Thus the line G08
relates to the $264+1655+\cdots+1152+161=38495$
families we've met earlier, with $m=8$ children, 
and with the 9 columns indicating the number of these with
$y=0,1,\ldots,8$ girls, and similarly for the other rows.
\end{small} 


{{\baselineskip12pt
\begin{small} 
\begin{verbatim}
       m      0       1      2      3      4     5     6     7    8    9   10  11 12 
G01 :  1 114609, 108719  
G02 :  2  47819,  89213, 42860 
G03 :  3  20540,  57179, 53789, 17395 
G04 :  4   8628,  31611, 44793, 28101,  7004 
G05 :  5   3666,  16340, 30175, 28630, 13740, 2839 
G06 :  6   1579,   7908, 17332, 22221, 15700, 6233, 1096 
G07 :  7    631,   3725,  9547, 14479, 13972, 8171, 2719,  436   
G08 :  8    264,   1655,  4948,  8498, 10263, 7603, 3951, 1152, 161 
G09 :  9     90,    713,  2418,  4757,  6436, 5917, 3895, 1776, 432,  66 
G10 : 10     30,    287,  1027,  2309,  3470, 3878, 3072, 1783, 722, 151,  30 
G11 : 11     24,     93,   492,  1077,  1801, 2310, 2161, 1540, 837, 275,  72,  8 
G12 : 12      7,     45,   181,   478,   829, 1112, 1343, 1033, 670, 286, 104, 24, 3  
\end{verbatim}
\end{small} 
}}


I note that for 1-child families,
the girl and boy frequencies are 0.487 and 0.513,
matching well what was found above for 8-children families. 

\begin{figure}[h]
\centering
\includegraphics[scale=\something]{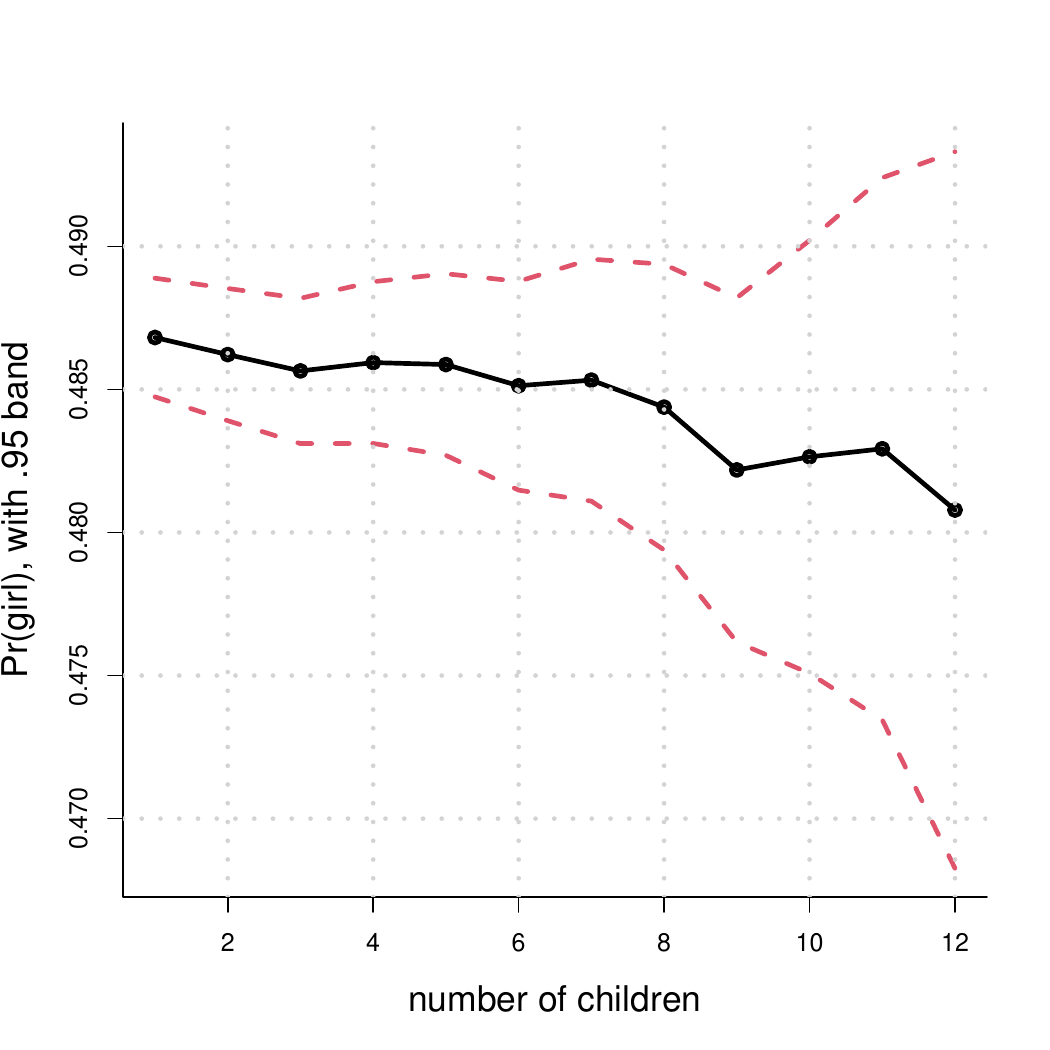}
\caption{The probability $\Pr(\girl)$, black line,
  with 95 \% confidence intervals, appears to go a bit
  down, as a function of family size, but the
  null hypothesis of $\Pr(\girl)$ staying the same is
  not siginificantly rejected.}
\label{figure:queen7}
\end{figure}

Using the methods above I can furthermore patiently
go through each family size, $m=2,\ldots,12$,
and estimate and assess $p_0$ and the extrabinomial
variation parameter $\sigma_0$ for each.
This leads first to Figure \ref{figure:queen7},
with the $\hatt p_0$, along with a 0.95 pointwise
confidence band. Yes, the $p=\Pr(\girl)$ 
appears to climb a little bit down, but this is not
significant. You may still fit your horizontal ruler
inside the band, and a proper test for the null hypotheses
that the twelve $p_m$ are identical does not signal
anything wrong, even with the enormous sample sizes involved.
So here I carefully disagree with claims made by
Edwards (1958) and Lindsey and Altham (1998).

\begin{figure}[h]
\centering
\includegraphics[scale=\something]{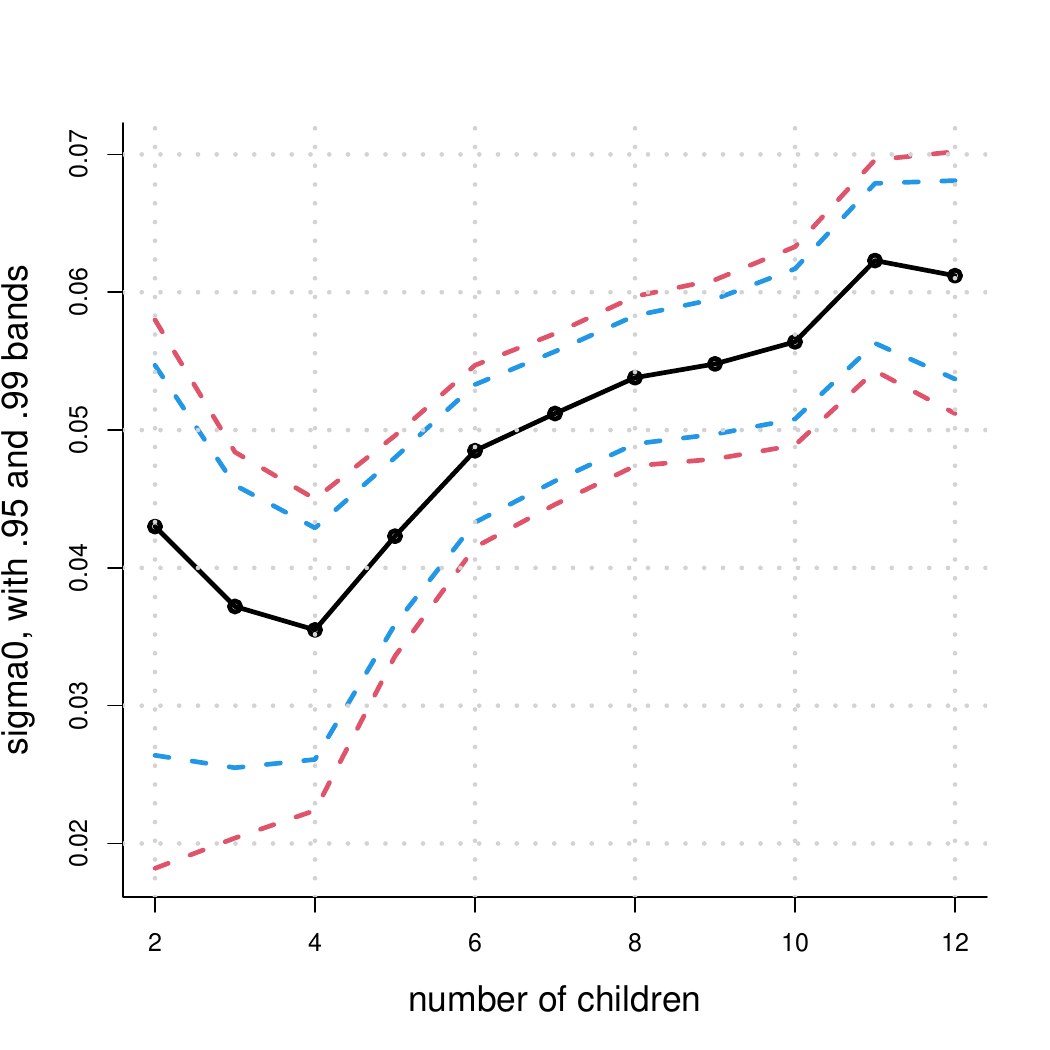}
\caption{Estimated overdispersion standard deviation $\sigma_0$,
  black points, inside 95 \% and 99 \% confidence intervals,
  as a function of family size, the number of children.}
\label{figure:queen8}
\end{figure}

What is however clear from these data is that the
level of extrabinomial dispersion increases with
the bigger families, as seen in Figure \ref{figure:queen8}, 
where I've used the confidence curves machinery
of Schweder and Hjort (2016) to calculate
0.95 and 0.99 intervals for $\sigma_0$, for each family size.
What we've learned above is that there might be several
competing reasons for this overdispersion. It could be that
the $p$ varies even more across families of size $m=10$
than for those of size $m=5$, etc., or there could be
dependencies in the chains of genders, and one might
think of yet more involved tiny mechanisms.
These would all be small (if they had been big,
humans studying humans would have noticed them much more
easily), but it is again part of this story of stories
that with such enormous sample sizes the statistical
microscopes can pick up even tiny effects and discrepancies
from what we might think of as Nature's default operations.

\section{Concluding comments}

{\it A: The quality of the Gei\ss ler (1889) data.}
Let me first point to an important detail I chose to pass
over above: when I discuss data for `families with 8 or more children',
etc., these actually have 9 or more children, with Gei\ss ler and
later writers having been careful enough to remove the gender
of the ninth child from his tables, in order not to disturb
the study of the human sex ratio via some present or imagined
`stopping rules'. So the tables above, with $y$ girls out of $m$ children,
pertain to the first $m$ children in families with at least $m+1$
children. 

Certain concerns have been voiced regarding the quality
and relative cleanliness of the Gei\ss ler data, which
apparently contain a certain but low number of repetitions,
etc. Generally the data quality is judged good enough
to yield good statistical analyses of the real gender
producing processes, however; cf.~Edwards (1958, 2005).
This is also my viewpoint (after having checked Edwards's
points), in particular for the work I've carried out
in this paper. 

{\it B: How many 8-children families must I check out before
  I spot the extrabinomial variance?}
I showed above that we need to check the gender of
some 10000 or ideally even 20000 children to really be
statistically fully certain that $p=\Pr(\girl)$ is not 0.50,
but a bit below 0.49. I've also commented that the
p-value associated with having computed the total-variance
to model-variance ratio $W_n=1.081$ test, for testing
the null hypothesis that the pure binomial is correct,
depends on the sample size, i.e. the number $n$ of
families with $m=8$ children in the dataset. With $n=1500$,
the p-value becomes very convincing, but not for e.g.~$n=500$. 

We may also ask how big $n$ ought to be before the statistical
power of the $W_n$ test is above level 0.95, say,
if the true state of affairs corresponds to the $\sigma_0=0.05$ 
(I estimated this extrabinomial standard deviation
parameter to be $\hatt\sigma_0=0.0538$ above). 
The figure below answers this question. It provides
the power of the $W_n$ test, calibrated to have significance
level 0.05, at that position in the alternative space
where $\sigma_0=0.05$, as a function of the number $n$ 
of families. I have done this via simulating $10^4$ 
realisations of $W_n$ at the null model, to identify
the rejection level, and then simulating another $10^4$ 
realisations of $W_n$ at the alternative position,
checking how often the test rejects the null.
We learn that about $n=4000$ families of size $m=8$ 
children are required for the careful statistician to be satisfied.


\begin{figure}[h]
\centering
\includegraphics[scale=\something]{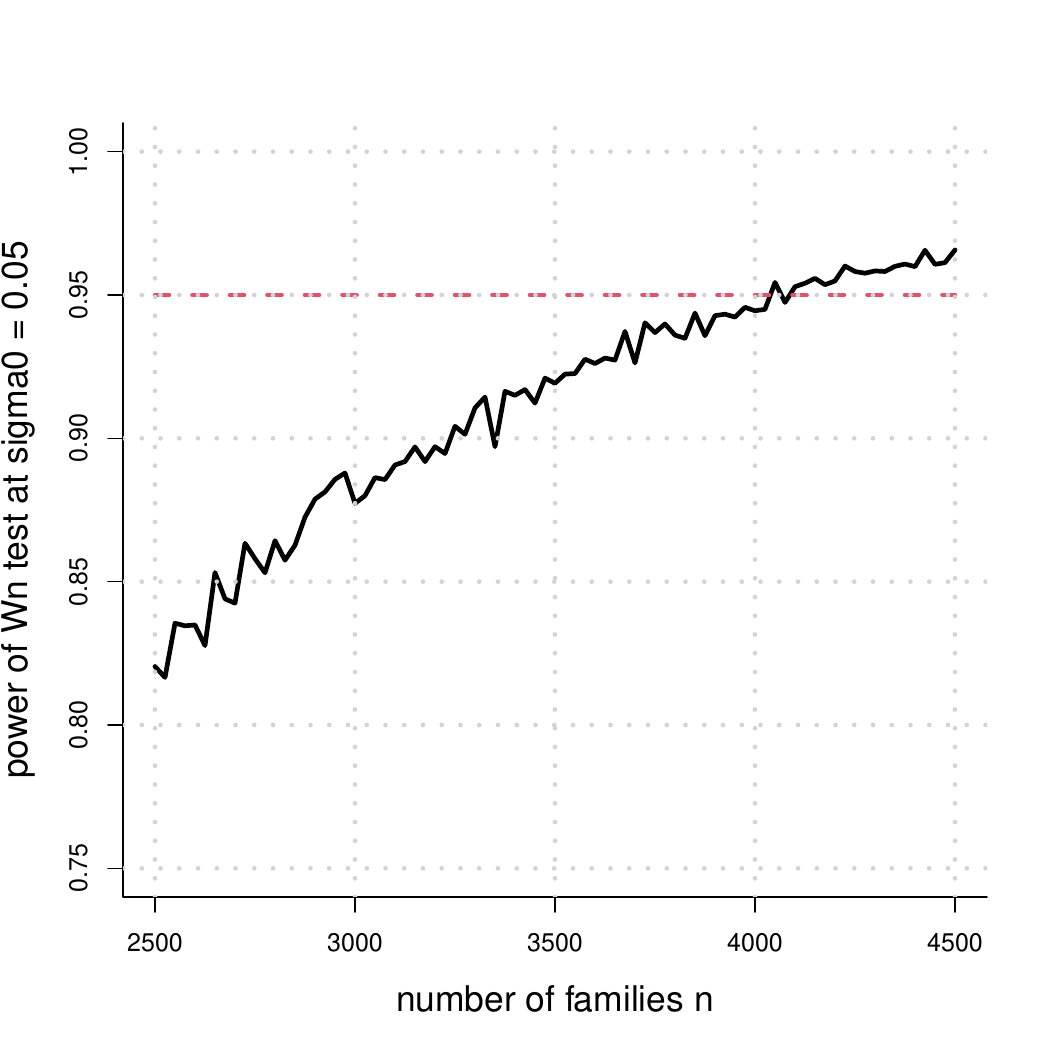} 
\caption{Power of the $W_n$ test, growing with the number
  $n$ of families. I needed about 4000 volunteer
  families of size 8 in order to be satisfied
  with my $W_n$ testing power.}
\label{figure:queen9}
\end{figure}

{\it C: Yet other models.}
Of course other models for the machineries of nature can
be put forward, fitted to these data, and compared;
see e.g.~Lindsey and Altham (1998) for a couple of
proposals. It would also be natural to use model
selection tools, as with the AIC, the BIC, the FIC
(cf.~Claeskens and Hjort, 2008, Hjort and Jullum, 2018),
to rank different models.

{\it D: The over-representation of all-girls and all-boys.}
So among the $n=38495$ families with $m=8$ children,
there were 264 with all-boys and 161 with all-girls.
These numbers are a bit bigger than what the binomial view
of the world would predict, namely 192 and 117.
With the beta-binomial model I can compute the theoretical
over-representation ratios for all-boys and for all-girls, as follows.

{{\baselineskip13pt
\begin{verbatim}
 2  1.007  1.008
 3  1.016  1.018
 4  1.029  1.032
 5  1.068  1.077
 6  1.137  1.155
 7  1.217  1.247
 8  1.320  1.379
 9  1.452  1.535
10  1.636  1.762
11  2.059  2.256
12  2.223  2.522 
\end{verbatim}
}}

\noindent 
The table gives the over-representation ratios for all-boys families,
$E_2(0)/E_1(0)$, and for all-girls, $E_2(m)/E_1(m)$, 
among families with $m$ children.

\begin{figure}[h]
\centering
\includegraphics[scale=\something]{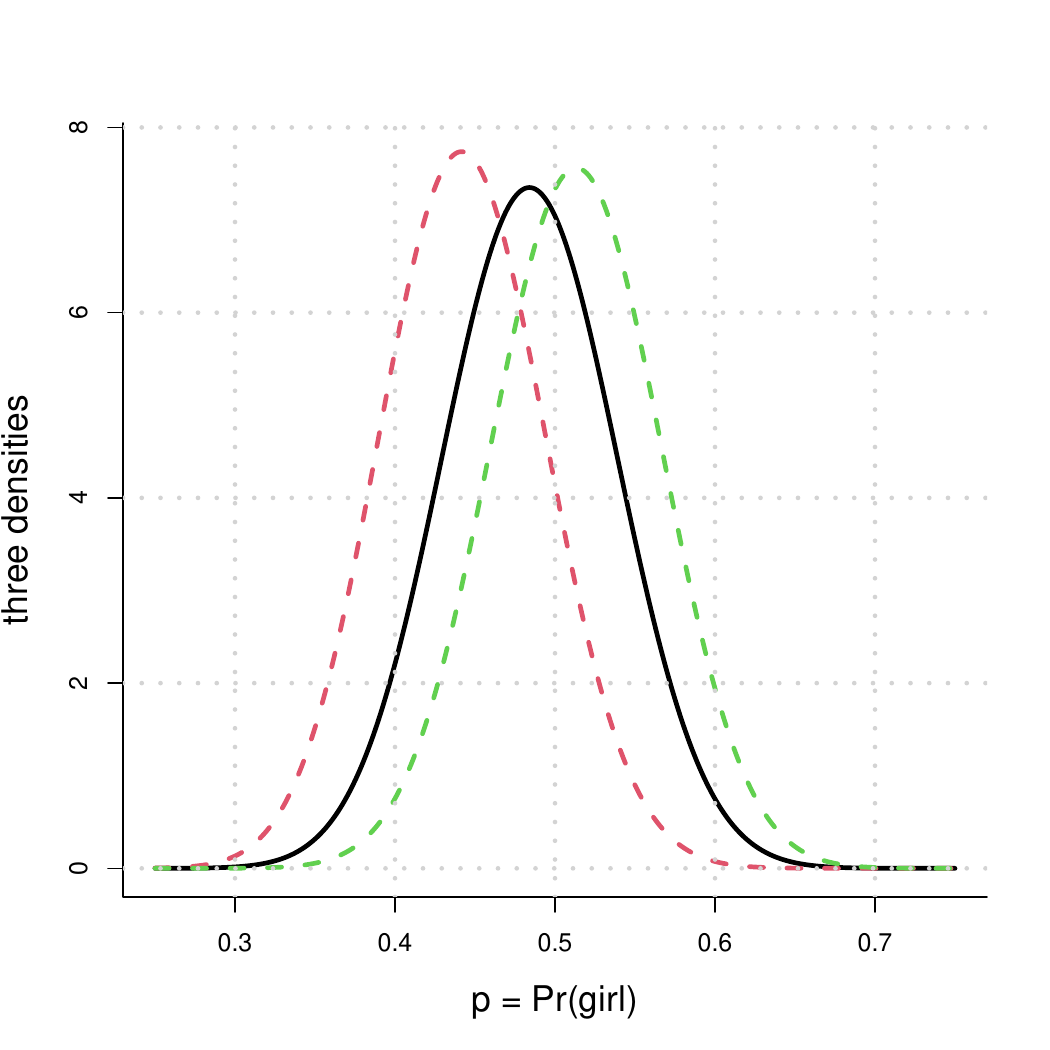} 
\caption{Beta densities for $p$ for three families:
  for the overall population (black, in the middle);
  for Kristin Lavransdatter (red, left);
  for a family with five sons (green, right).}
\label{figure:queen10}
\end{figure}

{\it E: Yet other data.}
Nichols (1905) discusses an interesting dataset
with about 3000 families, with 6 or more children,
mostly of Anglo-Saxon descent, from the 1600-1850 era.
His conclusions appear to match those obtainable
from Gei\aa ler (1889), regarding the basic probability
parameter $p=\Pr(\girl)$.

{\it F: Are more boys being born during wars?}
This question has been raised in the past, but as far
as I've understood it has been dismissed as a small
statistical fluke, with apparent bumps in some graphs
within reasonable stationary fluctuations after all.
However, Sir David Spiegelhalter (2015a, 2015b) comes
to the rescue, and finds that the old myth is not a myth,
but the Real Thing, complete with a reasonable
statistical argument (related to random visits
inside the menstrual cycles).

{\it G: Assessing p after all the births.}
The figure displays three Beta densities for $p=\Pr(\girl)$ 
The black in the middle is the overall one, valid in the
population at large, with parameters $(a,b)$ 
as estimated above. The red to the left is for Kristin Lavransdatter,
with parameters $(a,b+8)$; as we recall from reading
the Nobel Prize 1928 classic, she and Erlend Nikulausson
had eight sons: Naakve,
Bj\o rgulf, Gaute, Ivar, Skule, Lavrans, Munan, and Erlend.
The green to the right must be for Nils Lid and Dagny Fosse,
with parameters $(a+5,b)$, as per the intro photo.

{\it H: Ho Erna.}
``Jeg tror ikke jeg trenger å forklare hvordan dette gjøres.
Jeg skal heller ikke komme med noen pålegg.''
(Statsminister Solberg, nyttårstalen 2019.) 


\section*{References}

\begin{small}
\parindent0pt
\parskip3pt

Arbuthnot, J. (1710).
An argument for Divine Providence, taken from the constant regularity
observ'd in the births of both sexes.
{\it Philosophical Transactions},
vol.~27, 186--190.

Claeskens, G. and Hjort, N.L. (2008).
{\it Model Selection and Model Averaging.}
Cambridge University Press.

Edwards, A.W.F. (1958).
An analysis of Geissler's data on the human sex ratio.
{\it Annals of Human Genetics}, vol.~23, 6--15.

Edwards, A.W.F. (2005).
Sexes and statistics.
{\it Significance}, vol.~2, issue 4, 185--186.

Edwards, A.W.F. (2025).
{\it The Latin Square: Essays in defence of R.A.~Fisher.}
Cam Rivers Publishing, Cambridge, UK. 

Fisher, R.A. (1930).
{\it The Genetical Theory of Natural Selection.}
Clarendon Press, London. 

Geissler, A. (1878).
Über die Phantasmen während des Einschlafens.
{\it Philosophische Studien}, vol.~1, 83–93.

Geißler, A. (1889).
Beiträge zur Frage des Geschlechts verhältnisses der Geborenen.
{\it Zeit\-schrift des königlichen sächsischen statistischen Bureaus},
vol.~35, 1--24.

Hjort, N.L. (2016).
Recruitment Dynamics and Stock Variability:
  The Johan Hjort Symposium, some personal reflections.
{\it FocuStat Blog Post.} 

Hjort, N.L. (2019).
Your Mother is Alive with Probability One Half.
{\it FocuStat Blog Post.} 

Hjort, N.L. and Jullum, M. (2018).
Categorical model selection. Manuscript. 

Hjort, N.L. and Stoltenberg, E.Aa. (2026).
{\it Statistical Inference: 600 Exercises and 100 Stories.}
Cambridge University Press. 

Klotz, J. (1972).
Markov chain clustering of births by sex.
Proceedings of the Sixth Berkeley Symposium on Mathematical Statistics,
vol.~4, 173--185.

Klotz, J. (1973).
Statistical inference in Bernoulli trials with dependence.
{\it Annals of Statistics}, vol.~1, 373--379.

Lindsey, J.K. and Altham, P.M.E. (1998).
Analysis of the human sex ratio by using overdispersion models.
{\it Applied Statistics}, vol.~47, 149--157.

Nichols, J.B. (1905).
The sex-composition of human families.
{\it American Anthropologist (New Series)}, vol.~7, 24--36.

Schweder, T. and Hjort, N.L. (2016).
{\it Confidence, Likelihood, Probability:
  Statistical Inference With Confidence Distributions.}
Cambridge University Press, Cambridge.

Spiegelhalter, D. (2015a).
Sex-rated statistics. {\it Significance}, vol.~12, issue 4, 21--25.
\end{small}


\bigskip 

\begin{figure}[h]
\centering
\includegraphics[scale=0.44]{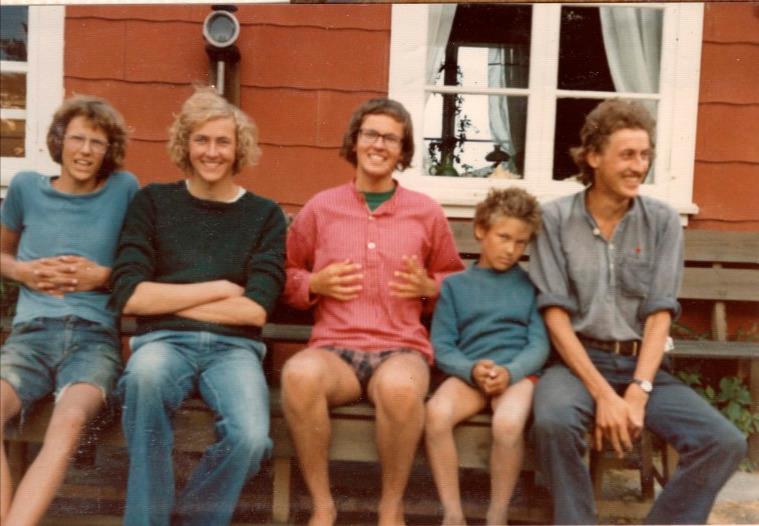}
\caption{$y=0$ girls, $m-y=5$ boys.}
\label{figure:queen11}
\end{figure}

\end{document}